# On Spectral and Energetic Characteristics of Erosional Plasma on the Basis of a Tin Alloy and of "Jumping Fireballs"

A.L. Pirozerski, A.I. Nedbai, V.A. Reznikov, E.L. Lebedeva and A.S. Khomutova

*Saint-Petersburg State University, Saint-Petersburg, 198504, Russia*

The ball lightning (BL) is one of the most enigmatic natural phenomena. Despite the long history of investigations its physical nature remains unknown. Experimental modeling of BL is one of the ways to resolve this problem. While up to now nobody was able to create real BL in laboratory in controlled and reproducible way, different researchers have developed several methods to generate long-living autonomous luminescent objects which resemble the natural phenomenon in some properties. One type of such objects is so-called "jumping fireballs" (JF), discovered and studied in [1], which are generated via impact of an erosional plasma jet, ejected from a polymethylmethacrylate capillary discharger, onto a tin anode. JFs had spherical shape with diameter of ~1-2 mm, glowed with bright yellow-reddish light, after falling on a surface they jumped up and down up to several tens times. Also JFs leaved aerogel tails which did not diffuse in the air but shrank into thin threads. Further studies of objects of the same class but generated without use of polymers were performed in [2].

In this paper we present the results of the spectral studies of erosive discharge with tin alloy electrodes and of the generated JFs, and experimentally determine internal energy of JFs using the calorimetric technique.

Simplified schematic diagram of the experimental setup is shown in Fig. 1a. The discharge circuit consists of a pulse storage capacitor ~1 mF x 5kV, an inductance L~0.25 mH integrated with a pulse ignition transformer, a protective discharge gap, main discharger and shunt resistor for measurements of the discharge current. The voltage on the main discharger was measured by a frequency-compensated divider. Oscillograms of the voltage and the current were registered using a Rigol DS2202A oscilloscope.

The main discharger included a dielectric plate with two adjustable dielectric supports carrying metallic guides for electrodes. Two piece of an industrial tin-copper alloy wire of length about 5 cm were used as the electrodes. The alloy composition was 97% Sn and 3% Cu. The electrodes were aligned opposite each other with a gap of 12 mm between their ends. The discharger was mounted at a height of about 30 cm over a support on which either a calorimeter described below or a vessel to collect the objects generated via the discharge were placed. Additionally, pyramidal or conic concentrators could be installed around or below the discharger.

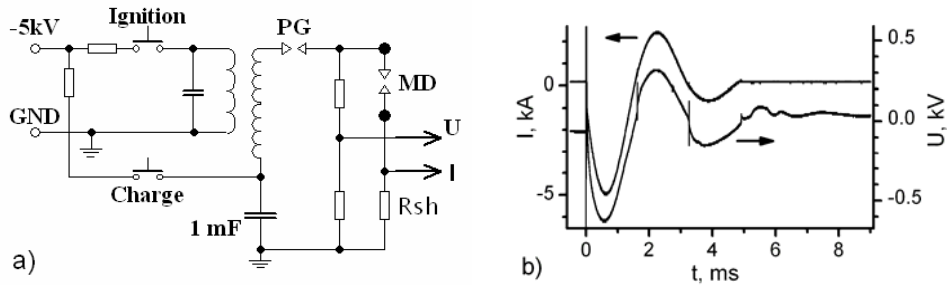

**FIGURE 1.** (a) Simplified schematic diagram of the experimental setup. PG – protective gap; MD – main discharger. (b) A typical oscillogram of the discharge current and the voltage on the main discharger.

Video recording of the experiments were performed via a Nikon 1 S1 digital camera. Spectra of the discharge and of JFs were recorded by a camcoder Sony HDR-HC9 using a modernized version of the original spectrograph [3]. The spectral range was 4000–6700 Å, with the maximal spectral resolution of ~1.5 Å. As a reference source a He-Hg lamp with a reflector was used. The spectral images were taken via the same spectrometer with the input slit removed.

To measure the energy of the jumping fireballs a simple calorimeter was constructed which consisted of a box made of 50 μm-thick copper foil surrounded by a thermal insulation. The temperature at several parts of the box was measured using a two junction copper-constantan thermocouple with the reference junction placed into melting ice. The calorimeter was calibrated with hot water.

As a result of the discharge a luminous plasmoid and a multitude of fast moving bright small point-like JFs are generated. The plasmoid fade away quickly and turns into a dust cloud. JFs fall on the laboratory table, some of them getting to the installed vessel or the calorimeter. The JFs leave bluish-gray thread-like tails consisting of tin oxide which are floating up in the atmosphere. The tails undergo a gradual turbulization, break up into smaller pieces and shrink. After falling on surface, JFs are actively moving on it either gliding or jumping to height up to few centimeters. Structure of the generated aerogel threads on the scales from hundreds microns to tens nanometers, and interaction of JFs with different targets were discussed elsewhere [2].

An oscillogram taken at a typical experiment is shown in Fig. 1b. The initial voltage of the capacitive storage was 3.3 kV. As it may be seen from the figure, the main stage of the discharge include three half-period of oscillations and has the duration of ~4.9 ms with maximal absolute magnitude of the current being ~5 kA. The fact that voltage on the discharger did not return to zero after this time means that a weak current flows in the discharge circuit for a longer time. Final voltage of the capacitive storage was usually of the opposite polarity that the initial one. Integration of the oscillogram gives the total energy input to the discharge at its main stage to be of 2.99 kJ, which, at the decrease of the electrode mass at the discharge being of 0.186 g, means about 19.8 eV/at. It is an upper estimate, in particular, due to energy losses for heating remaining parts of the electrodes and constructional materials of the discharger).

The discharge spectrum in visible spectral range is shown in Fig.2. Several lines are present superimposed on a sufficiently bright continuous spectrum. Most of noticeable line marked in Fig.2 was identified using [4-6] and found to belong to tin and copper atoms and first ions, see Table 1, in accordance with the alloy composition. Note that lines of Cu II and Sn II have the excitation energies of about 15-17 eV and 9-11 eV, respectively. Taking into account that ionization energies for copper and tin are 7.72 eV and 7.34 eV, respectively, this implies the presence in the erosive discharge plasma of excited states with energies up to 25 eV. Taking into account that the discharge occurs in the ambient conditions this is very unusual. For example, in spectroscopic studies of the interaction of the capillary discharge jet with the metal foils [7] no lines of atomic ions were observed. To elucidate mechanisms responsible for the formation of these excited states further investigations are required.

Because the luminescence of the JFs themselves is sufficiently lower, it was difficult to obtain their spectra in normal mode, i.e. with the input slit installed.

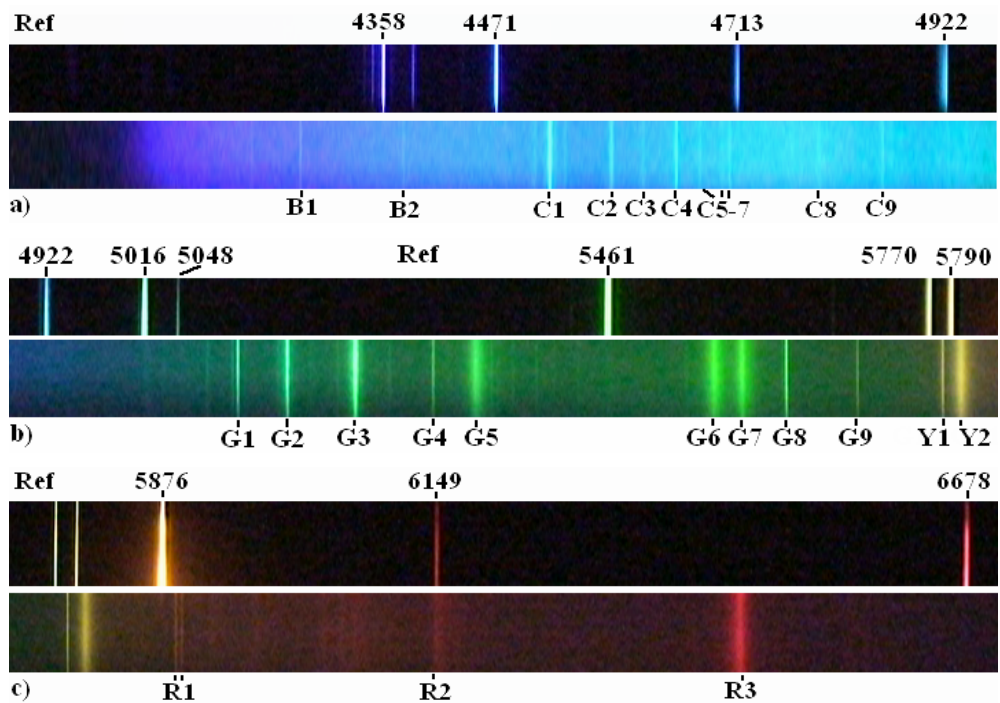

**FIGURE 2.** The discharge spectra in violet-blue (a), blue-yellow (b) and yellow-red (c) spectral ranges. Upper images (marked as Ref) are reference spectra of the He-Hg lamp installed behind the discharger, lower images are the discharge spectra. The wavelengths are given in angstroms. Noticeable lines or cants are marked by capitals letters with numbers in the bottom of each image.

**TABLE 1.** Noticeable lines in the spectrum of the erosive discharge with the tin alloy electrodes.

| Line | Ion/atom | Wavelength, Å | Excitation energy, eV | Line | Ion/atom | Wavelength, Å | Excitation energy, eV |
|---|---|---|---|---|---|---|---|
| B1 | Cu I | 4275.13 | 7.74 | G4 | Cu I | 5292.5 | 7.74 |
| B2 | Cu I | 4378.2 | 7.8 | G5 | Cu II | 5331.63 | 16.0 |
|    |      |         |     |    | Sn II | 5332.3 | 11.19 |
| C1 | Sn I | 4524.74 | 4.87 | G6 | Sn II | 5561.9 | 11.20 |
|    |      |         |      |    | Cu II | 5560.6 | 17.22 |
|    |      |         |      |    | Cu II | 5562.07 | 16.85 |
|    |      |         |      |    | Cu II | 5562.64 | 16.66 |
| C2 | Cu I | 4586.95 | 7.8 | G7 | Sn II | 5588.81 | 11.07 |
| C4 | Cu I | 4651.13 | 7.74 | G8 | Sn I | 5631.71 | 4.33 |
| C5 | Cu I | 4674.76 | 7.8 | G9 | Cu I | 5700.24 | 3.82 |
| C6 | Cu I | 4698.7 | 7.88 | Y1 | Cu I | 5782.13 | 3.79 |
| C7 | Cu I | 4704.6 | 7.74 | Y2 | Sn II | 5796.66 | 11.07 |
|    |      |        |      |    | Sn II | 5798.86 | 11.07 |
|    |      |        |      |    | Cu II | 5801.13 | 15.53 |
| G1 | Cu I | 5105.54 | 3.82 | R1 | Cu II | 5890.4 | 16.75 |
|    |      |         |      |    | Cu II | 5897.97 | 16.56 |
| G2 | Cu I | 5153.24 | 6.19 | R2 | Cu I | 6147.31 | 16.56 |
| G3 | Cu I | 5218.20 5220.7 | 6.19 6.19 | R3 | Sn II | 6453.58 | 8.97 |

Hence, taking into account the small size of the JFs, the slit was removed and the spectral images were registered instead of spectra, see Fig. 3. At that, a conic concentrator was used to make JFs be closer to the normal slit position. It should be noted that despite of a blur due to both the finite size and the motion of JFs, and also a distortion of the wavelength calibration due to offset of the mean position of JFs, spectral images reproduce to some extent the features of the spectrum. As seen from Fig. 3, the spectral images show no signs of any spectral lines or cants, so we assume that JFs have continuous spectra.

For direct measurements of the JF energy calorimetric studies were carried out. The calculated idealized sensitivity of the calorimeter was ~0.46 K/J. However, the calibration of the calorimeter using hot water showed that the heat recovery coefficient (i.e., the ratio of the energy evaluated as maximal calorimeter temperature increase divided by the idealized sensitivity to real value of the energy) was only ~0.15.

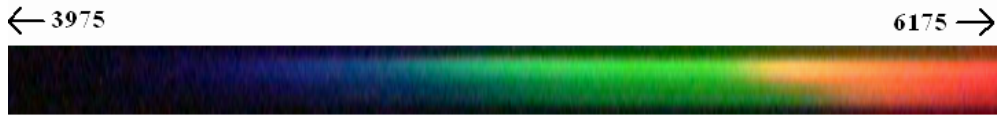

**FIGURE 3.** The spectral image of a jumping fireball. The wavlengthes (in angstroms) corresponds to the left and right edges of the corresponding spectrum which would be taken with the input slit installed.

Mass of the JFs hitting upon the calorimeter found by weighing was of order of milligrams or tens milligrams. Calculated specific internal energy of the JFs (taking into account the heat recovery coefficient) varies from ~1 to 7 eV/at. The higher values correspond to the greater JF masses and, therefore, seem to be more reliable. Note, that the specific energy required to heat the tin from room temperature to the melting one (505.08 K) is ~0.06 eV/at, the specific melting heat is ~0.07 eV/at, the specific energy to heat the melt from melting temperature to the boiling one (2875 K) is >0.7 eV/at, the specific energy of the vaporization is ~3.1 eV/at. The sum of all this energies is >4 eV/at. The enthalpy of formation of tin oxide (IV) from solid tin and gaseous oxygen is ~6 eV/at. Taking into account uncertainty of the heat recovery coefficient and its possible dependence on mass of the objects and places where they hit into the calorimeter box, the present results can not give unambiguous answer whether the available chemical and thermal energy of JFs is sufficient to account for the result of the calorimetric measurements or a fraction of the internal energy of JFs is stored in some other form.

In conclusion, the present studies show the presence of excited ions with excitation energies of order 20-25 eV in the plasma of the erosional discharge with tin alloy electrodes. Luminescence of JFs has continuous spectrum, and its internal energy, determined via calorimetric studies, is of the same order of magnitude as their available thermal and chemical energy, however the presence of other forms of energy storage cannot be excluded definitely.